\documentclass[12pt,preprint]{aastex}

\shorttitle{}
\shortauthors{Nesvorn\'y et al.}

\begin{document}
\baselineskip 19.pt

\title{Formation of Kuiper Belt Binaries by Gravitational Collapse}

\author{David Nesvorn\'y$^1$, Andrew N. Youdin$^2$, Derek C. Richardson$^3$}
\affil{(1) Department of Space Studies, Southwest Research Institute,\\
1050 Walnut St., Suite 300, Boulder, CO, 80302, USA}
\affil{(2) Canadian Institute for Theoretical Astrophysics, University of Toronto,\\
60 St. George St., Toronto, ON, M5S 3H8, Canada}
\affil{(3) Department of Astronomy, University of Maryland,\\ College Park, MD, 
20742-2421, USA}

\begin{abstract}
A large fraction of $\sim$100-km-class low-inclination objects in the classical Kuiper Belt (KB) are 
binaries with comparable mass and wide separation of components. A favored model for their formation 
was capture during the coagulation growth of bodies in the early KB. Instead, recent studies 
suggested that large, $\gtrsim$100-km objects can rapidly form in the protoplanetary disks when 
swarms of locally concentrated solids collapse under their own gravity. Here we examine the possibility 
that KB binaries formed during gravitational collapse when the excess of angular momentum prevented the 
agglomeration of available mass into a solitary object. We find that this new mechanism provides a 
robust path toward the formation of KB binaries with observed properties, and can explain wide 
systems such as 2001 QW$_{322}$ and multiples such as (47171) 1999 TC$_{36}$. Notably, the 
gravitational collapse is capable of producing $\sim$100\% binary fraction for a wide range of the swarm's 
initial angular momentum values. The binary components have similar masses ($\sim$80\% have the secondary-over-primary
radius ratio $>$0.7) and their separation ranges from $\sim$1,000 to $\sim$100,000 km. The binary orbits have 
eccentricities from $e=0$ to $\sim$1, with the majority having $e<0.6$. The binary orbit inclinations 
with respect to the initial angular momentum of the swarm range from $i=0$ to $\sim$90$^\circ$, with most cases 
having $i<50^\circ$. The total binary mass represents a characteristic fraction of the collapsing swarm's 
total initial mass, $M_{\rm tot}$, suggesting $M_{\rm tot}$ equivalent to that of a radius $\sim$100 to 250-km 
compact object. Our binary formation mechanism also implies that the primary and secondary components 
in each binary pair should have identical bulk composition, which is consistent with the current 
photometric data. We discuss the applicability of our results to the Pluto-Charon, Orcus-Vanth, 
(617) Patroclus-Menoetius and (90) Antiope binary systems. 

\end{abstract}


\section{Introduction}

The existence of binary Kuiper Belt Objects (KBOs) other than Pluto-Charon (Christy \& Harrington 1978) 
has been suspected since the discovery of the Kuiper belt (Jewitt \& Luu 1993), but it 
was not until December 2000 that the first binary KBO, 1998 WW$_{31}$, was detected by direct 
ground-based imaging (Veillet et al. 2001, 2002). Recent observations indicate that $\sim$30\% of 
100-km-class classical cold KBOs with orbital inclinations $i<5^\circ$ are binaries (Noll et al. 2008a,b; 
$>0.06$ arcsec separation, $<$2 mag magnitude contrast). Binaries with larger primaries, large magnitude 
differences and smaller separations may be even more common (Brown et al. 2006, Weaver et al. 2006) and 
probably require a different formation mechanism (e.g., Canup 2005).

The properties of known binary KBOs differ markedly from those of the main-belt and near-Earth asteroid 
binaries (Merline et al. 2002, Noll et al. 2008a). The 100-km-class binary KBOs identified so far 
are widely separated and their components are similar in size. These properties defy standard ideas 
about processes of binary formation involving collisional and rotational disruption, debris re-accretion, 
and tidal evolution of satellite orbits (Stevenson et al. 1986). They suggest that most binary KBOs 
may be remnants from the earliest days of the Solar System. If so, we can study them to learn about 
the physical conditions that existed in the trans-Neptunian disk when large KBOs formed.    

The Kuiper belt (Kuiper 1951, Jewitt \& Luu 1993) provides an important constraint on planet formation. 
To explain its present structure, including the large binary fraction among the classical cold KBOs 
discussed above, we must show how the 100-km-size and larger bodies accreted from 
smaller constituents of the primordial trans-Neptunian disk. Two main possibilities exist: (1) 
Hierarchical Coagulation (hereafter HC), where two-body collisions between objects in a dynamically cold 
planetesimal disk lead to objects' accretion and growth; and (2) Gravitational Instability (hereafter GI), 
where the gas-particle effects and/or gravitational instabilities produce concentrations of gravitationally 
bound solids followed by their rapid collapse into large objects. We briefly comment on these 
theories below.

As for HC, Stern (1996), Stern \& Colwell (1997), Kenyon \& Luu (1998, 1999), and Kenyon 
(2002) conducted simulations of the primordial `bottom-up' process involving collisional
accumulation of small KBOs into larger ones (also see Kenyon et al. 2008 for a review). 
They found that two competing physical processes, growth by mergers and erosion by fragmentation,
determine the final result. According to these studies, the observed KBOs can only form by HC in 
$\lesssim 10^8$ years if: (i) the orbits in the belt were initially much more circular and planar 
than they are now ($e \sim i \sim 10^{-4}$--$10^{-2}$ compared to present eccentricities $e \sim 0.1$ 
and inclinations $i \sim 10^\circ$), and (ii) the initial disk mass was $\sim$100--1000 
times larger than the current KB mass, $M_{\rm KB}\sim0.01$-0.1 $M_{\rm Earth}$ (Trujillo et al. 2001a,b, 
Gladman et al. 2001b, Bernstein et al. 2004, Fraser et al. 2008). 

%ANY: Revised GI paragraph
The GI hypothesis has been advanced by recent breakthroughs in theory and simulation (see Chiang \& Youdin 
(2010) for a review).  The classical GI of a particle-rich nebula mid-plane 
(Safronov 1969; Goldreich \& Ward 1973; Youdin \& Shu 2002) can be prevented by even a modest 
amount of stirring from a turbulent gas disk (Weidenschilling 1980; Cuzzi et al. 1993). However, 
particles can also clump in a turbulent flow (e.g., Cuzzi et al. 2001, 2008; Johansen et al. 2006).  
The streaming instability (Youdin \& Goodman 2005) is a powerful concentration mechanism by which weak 
particle clumps perturb the gas flow in a way that increases their amplitude (Youdin \& Johansen 2007, 
Johansen \& Youdin 2007).  Simulations of rocks in a gas disk find that streaming instability-induced  
clumping produces gravitationally-bound clusters of solids, either with (Johansen et al. 2007) or without 
(Johansen et al. 2009) large scale MHD turbulence.   These clumps exceed the mass of compact $100$ km radius 
planetesimals.  The local disk metallicity (relevant for the amount of condensed solids) needs to slightly 
exceed Solar abundances in order to counter turbulent stirring and trigger strong clumping (Youdin \& Shu 
2002, Johansen et al. 2009).  Much work remains to determine the relative roles of GI and HC in 
the Solar System and beyond.

%ANY: Previous GI paragraph 
%As for GI, much work has yet to be done. The classical GI of a particle-rich nebula mid-plane 
%(Safronov 1969; Goldreich \& Ward 1973; Youdin \& Shu 2002) can be prevented by even a modest 
%amount of stirring from a turbulent gas disk (Weidenschilling 1980; Cuzzi et al. 1993). However, 
%particles can also clump in the turbulent flow (e.g., Cuzzi et al. 2001, 2008; Johansen et al. 2006; 
%see Chiang \& Youdin 2009 for a review). For example, Johansen et al. (2007) conducted simulations 
%of self-gravitating boulders embedded in a gas disk with MHD turbulence. They showed that the conditions 
%required for the GI can be locally achieved when pressure gradients concentrate boulders in turbulent 
%eddies. Interestlingly, even in the absence of large-scale turbulence, important particle concentrations 
%can be induced by the streaming instability assuming that the protoplanetary disk's metalicity was 
%slightly larger than the solar value (Youdin \& Goodman 2005, Johansen et al. 2009).

\subsection{Binary Formation in HC}

Several theories have been proposed for the formation of binary KBOs in the HC model: 
{\bf (i)} Gravitational reactions during encounters among {\it three} KBOs may redistribute 
their kinetic energy enough so that two KBOs end up in a bound orbit, forming a binary,
with the third object carrying away the excess energy (Goldreich et al. 2002). 
{\bf (ii)} An encounter between {\it two} KBOs can lead to binary formation provided that 
the encounter energy is dissipated by some mechanism. Goldreich et al. (2002) proposed that 
in the early KB, the energy dissipation occurred due to the effects of dynamical friction 
(Chandrasekhar  1943, Binney \& Tremaine 1987) from numerous small bodies passing through 
the encounter zone (also see Schlichting \& Sari 2008a,b). 
{\bf (iii)} The collisional merger of two bodies within the sphere of influence of a third 
body can also produce a binary. Such mergers could have been a common occurrence in the early 
KB (Weidenschilling 2002).
{\bf (iv)} Physical collisions invoked in (iii) can produce close binaries with a large 
primary-to-satellite mass ratio. Subsequent scattering encounters with large KBOs can
cause exchange reactions in which the small satellite is replaced by a larger and 
more distant secondary (Funato et al. 2004). 
{\bf (v)} A transitory binary system may form by chaos-assisted temporary capture.\footnote{The 
chaos-assisted temporary capture is an important feature of 3-body dynamics. It occurs when two
bodies are trapped into a thin region between stable-bound and unbound energy states, where orbits 
are chaotic but confined by phase space constraints. In absence of dissipation, the two captured 
bodies would temporarily orbit each other, as if in a wide binary system, before separating after 
typically only a few periods.} 
The binary can then be stabilized by a sequence of discrete encounters with small 
background planetesimals (Astakhov et al. 2005; Lee et al. 2007). This model invokes a different variant of capture 
than model (ii) but uses encounters with small bodies as in (ii) to shrink and stabilize the 
binary orbit. 

Some of the models listed above seem to be too inefficient to explain the observed high binary 
fraction and/or do not match other constraints. For example, according to Goldreich et al. (2002), collisionless 
gravitational interactions are more efficient in forming the observed, widely separated binaries 
in the primordial KB than (iii). Also, model (iv) leads to binary eccentricities $e \gtrsim0.8$ and very large 
semimajor axes, while observations of binary KBOs indicate moderate eccentricities and semimajor axes 
that are only a few percent of the Hill radius (Noll et al. 2008a, Grundy et al. 2009), except for 1998 WW$_{31}$ 
with $e=0.82$ (Veillet et al. 2002) and 2001 QW$_{322}$ with $a=120,000$ km (Petit et al. 2008). 

Schlichting \& Sari (2008a) estimated that chaotic capture in (v) should be less common than direct 
capture in (i) or (ii). Both (i) and (ii), however, put rather extreme requirements on
the size distribution of objects in the primordial trans-planetary disk (Goldreich et al. 2002).
Specifically, the encounter speeds between the 100-km-class KBOs, $V_{\rm enc}$, need to be similar to or 
preferably lower than the Hill speed, $V_{\rm enc} \lesssim V_{\rm Hill} = \Omega_{\rm Kep} R_{\rm Hill} 
\sim 0.2$~m~s$^{-2}$. Here, $\Omega_{\rm Kep} $ denotes the orbital frequency of a Keplerian orbit with 
semimajor axis $a$, $R_{\rm Hill} = a (M/3M_{\rm Sun})^{1/3}$ is the Hill sphere of a body with mass 
$M$, $M_{\rm Sun}$ is the mass of the Sun, and the above numeric value was given for $a=30$ AU and mass 
corresponding to a 100-km-diameter 
sphere with 1~g~cm$^{-3}$ density. To satisfy this condition, Goldreich et al. postulated an initially 
bimodal size distribution of planetesimals in the primordial disk with $\sigma/\Sigma\sim10^3$, where 
$\sigma$ and $\Sigma$ are the surface densities of small and 100-km-class bodies, respectively. The effects 
of dynamical friction from the very massive population of small bodies can then indeed ensure that 
$V_{\rm enc} \lesssim V_{\rm Hill}$ long enough for binary formation to occur. 

It is not clear whether the bimodal size distribution with $\sigma/\Sigma\sim10^3$ actually occurred in the 
early KB. The binary formation rates in (i) and (ii) are apparently almost a step function in $\sigma/\Sigma$ 
with values $\sigma/\Sigma<5\times10^2$ leading to only a small fraction of binaries in the population. 
In addition, mechanism (ii) that is expected to prevail over (i) for $V_{\rm enc}<V_{\rm Hill}$ produces retrograde 
binary orbits (Schlichting \& Sari 2008b), while current observations indicate a more equal mix of prograde and 
retrograde orbits (Noll et al. 2008a, Petit et al. 2008, Grundy et al. 2009). This could suggest that binary KBOs 
formed by (i) when $V_{\rm enc} \sim V_{\rm Hill}$ (Schlichting \& Sari 2008b) and, inconveniently, implies a very narrow 
range of $\sigma/\Sigma$. 
    
\subsection{New Model for Binary Formation in GI}  

Benecchi et al. (2009) reported resolved photometric observations of the primary and secondary
components of 23 binary KBOs. They found that the primary and secondary components of each observed 
binary pair have identical colors to within the measurement uncertainties. On the other hand, the 
wide color range of binary KBOs as a group is apparently indistinguishable from that of the population of single 
KBOs. These results can be difficult to understand in (some of) the models of binary KBO formation discussed in
1.1. Instead, the most natural explanation is that binary KBOs represent snapshots of the local composition 
mix in a nebula with important temporal and/or spatial gradients.\footnote{Note that ejecta exchange (Stern 2008) in 
comparable mass binaries would produce a color distribution with a smaller variance by ``averaging'' 
the component colors, which is not observed (Benecchi et al. 2009).}  

\begin{figure}
\epsscale{0.45}
\plotone{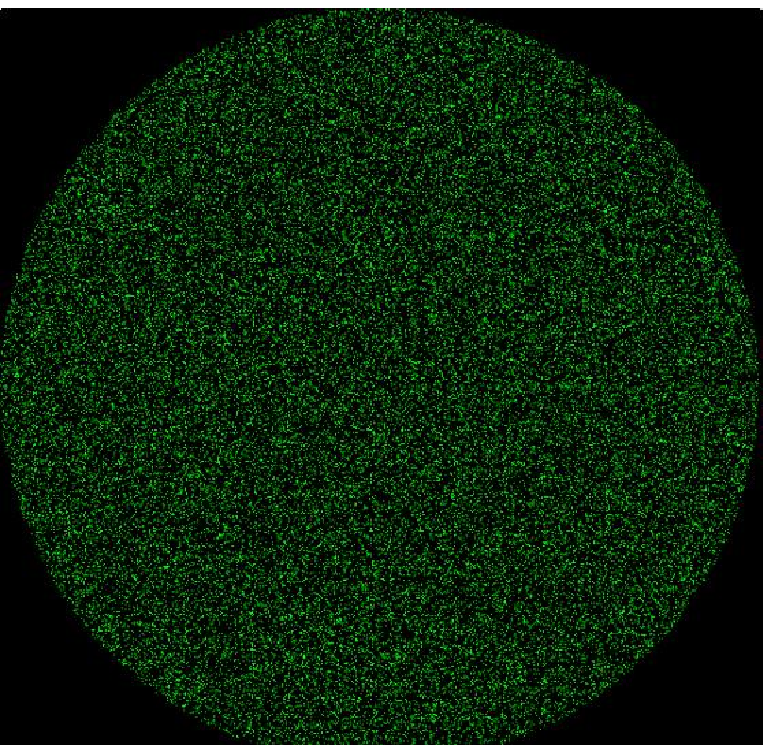}\hspace*{5.mm}
\plotone{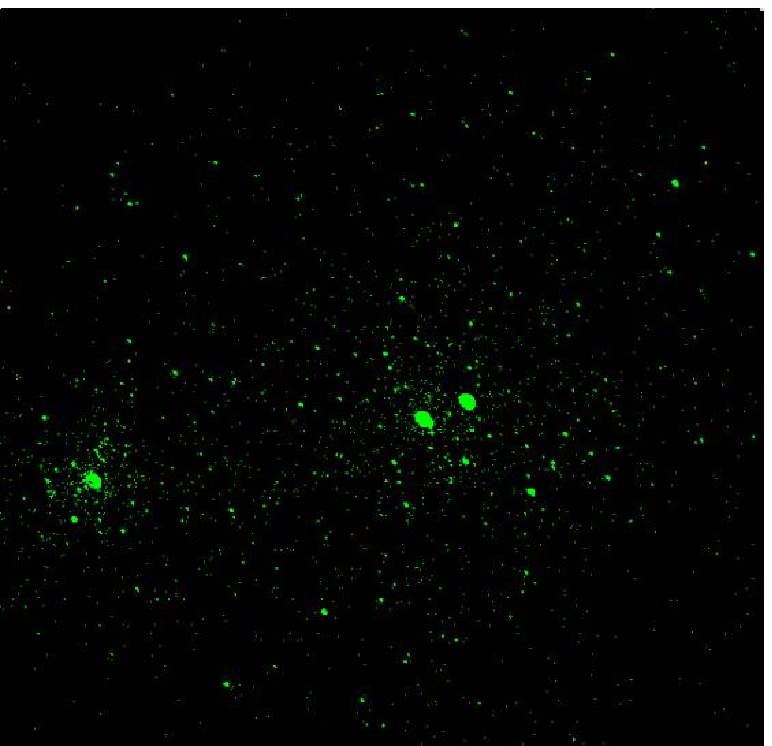}
\caption{Illustration of binary formation by gravitational collapse. Here we tracked $2.5\times10^5$ 
particles as they evolve by gravitational interactions and mutual inelastic collisions. To start with, 
we distributed the particles in a spherical volume $2\times10^5$~km across and gave them small random 
velocities (left panel). Slow initial rotation was given to the swarm to mimic the motion induced 
from the background turbulence and/or other processes. After collapse, a temporary triple system
formed with nearly equal 100-km size components (right panel). Subsequent ejection of the outer 
component or a collision of the inner pair left behind a binary system with $\sim$$10^4$-$10^5$~km 
separation. Figure from Nesvorn\'y (2008). Size of objects scaled for visibility.}
\label{n08}
\end{figure}

The observed color distribution of binary KBOs can be easily understood if KBOs formed by GI. The common 
element invoked by various GI models is the final stage of gravitational collapse when the gravitationally 
bound pebbles and boulders are brought together, collide and eventually accrete into large objects. We 
envision a situation in which the excess of angular momentum in a gravitationally collapsing swarm prevents 
formation of a solitary object. Instead, a binary with large specific angular momentum forms from 
local solids, implying identical composition (and colors) of the binary components.
Moreover, binaries with similarly sized components are expected to form in this model, because similar 
components maximize the use of the collapsing cloud's angular momentum (Fig. \ref{n08}; Nesvorn\'y 2008).\footnote{The 
orbital angular momentum of binary components increases with their mass ratio, $q \leq 1$, as $q/(1+q)^2$ for fixed 
semimajor axis, eccentricity and total mass.} 

Our model for binary KBO formation is similar to that of binary stars from the collapse of a rotating 
molecular cloud core (Kratter et al. 2008), and more specifically to binary star formation in fragmenting 
disks around black holes (e.g., Alexander et al. 2008). It has not been studied in the context of planetary 
science. For example, while Johansen et al. (2007, 2009) investigated the formation of  gravitationally bound 
concentrations of solids, they did not follow the final stage of gravitational collapse in detail because 
the spatial resolution of their code was limited by the need to resolve much larger scales of disk dynamics.

Here we conduct $N$-body numerical simulations of a gravitationally collapsing segment of disk solids to 
determine whether the observed 100-km-class binary KBO could have formed in the GI model. We attempt to 
``reverse engineer" the conditions that give rise to binary formation by varying the initial set of parameters.  
This is because precise initial conditions in a bound clump are uncertain due to the complex physics of 
particles in turbulent accretion disks. We do not attempt to extract our initial data from the Johansen
et al. (2007, 2009) simulations because they have low resolution ($<$10 grid cells) across the densest
clumps. We describe our integration method and setup in section 2. The results are presented in section 3 and 
discussed in section 4.

\section{Method}

Our simulations of gravitational collapse were performed with a modified version of the $N$-body cosmological
code PKDGRAV (Stadel 2001), described in Richardson et al. (2000) (also see Leinhardt et al. 2000, 
Leinhardt \& Richardson 2002). PKDGRAV is a scalable, parallel tree code 
that is the fastest code available to us for this type of simulation. A unique feature of the code is the 
ability to rapidly detect and treat collisions between particles.  We used $N=10^5$ particles per run.
Each PKDGRAV particle was given initial mass $M=M_{\rm tot}/N$, where $M_{\rm tot}$ was the assumed total 
mass of the gravitationally unstable swarm. Initially, the PKDGRAV particles were distributed in a spherical 
volume with radius $R_{\rm tot} < R_{\rm Hill}=(GM_{\rm tot}/3 \Omega_{\rm Kep}^2)^{1/3}$, in which 
self-gravity dominates ($G$ is the gravitational constant).

The initial velocities of PKDGRAV particles were set to model the collapse phase that occurs after some GI. 
Since the exact GI conditions are uncertain due to the modest resolution and uncertainties in the existing 
instability calculations, we sampled around a range of the initial velocities to see how different assumptions 
would influence the results. Specifically, we gave the swarm uniform rotation with several different values 
of $\Omega \lesssim \Omega_{\rm circ}$, where $\Omega_{\rm circ} = V_{\rm circ}/R_{\rm tot}$ and 
$V_{\rm circ}=\sqrt{GM_{\rm tot}/R_{\rm tot}}$ is the speed of a particle in a circular orbit about the cloud 
at $R_{\rm tot}$. In addition, particles were also given random velocities with characteristic 
speed $V_{\rm rand}<V_{\rm circ}$. 

The Keplerian shear was included in the Hill approximation as in Tanga et al. (2004) (except that no periodic 
boundaries were imposed). We also conducted experiments where the Sun was directly included in the simulations 
as a massive PKDGRAV particle. The results obtained with these two methods were similar. Since 
$\Omega_{\rm circ}/\Omega_{\rm Kep} = \sqrt{3} (R_{\rm Hill}/R_{\rm tot})^{3/2}$, the shearing effects 
quickly diminish for $R_{\rm tot} < R_{\rm Hill}$, because the cloud is initially compact and collapses in a 
fraction of the orbital period. 

Given the exploratory nature of our investigation, we neglected certain physical ingredients that should be 
less significant, but could be added to the next generation of models. Specifically, gas drag was ignored 
because our estimates show that the effects of gas drag should be small relative to collisional damping inside 
the gravitationally bound clump (appendix A). In addition, `mass loading' (see, e.g., Hogan \& Cuzzi 2007) damps 
turbulence inside dense particle clumps, making it safe to ignore the forcing of particle motions by the turbulent 
gas.  The evolution of particle speeds in our simulations is set by gravitational interactions and 
physical collisions, the dominant effects during the final stage of collapse.

We ignored collisional fragmentation of bodies in the collapsing swarm because the expected 
impact speeds are low (see section 3) and we can develop a better understanding of the collapse process with
simple models. Note that debris produced by disruptive collisions between bodies 
in the collapsing swarm are gravitationally 
bound so that even if fragmentations occur, fragments are not lost. The fragmentation can be included in 
the next generation models using scaling laws developed for low-speed collisions between icy bodies (e.g., Leinhardt 
\& Stewart 2009; Stewart \& Leinhardt 2009), even though it can be challenging to deal with the full complexity 
of the collisional cascade. 

We divided the integrations into two suites. 
In the first suite of our `core' simulations, we used a simple physical model of collapse and covered a regular 
grid of parameter values in $M_{\rm tot}$, $\Omega$ and $r$. Specifically, we used $\Omega=0.5$, 0.75, 1.0 and 
1.25$ \Omega_{\rm circ}$ and $R_{\rm eq} =100$, 250 and 750 km, where $R_{\rm eq}$ is the equivalent 
radius of a sphere with mass $M_{\rm tot}$ and $\rho=1$~g~cm$^{-3}$. The collisions between PKDGRAV particles 
were treated as ideal mergers.\footnote{In this approximation, every collision resulted in a merger, with no mass
loss, and the resulting body was a single sphere of mass equal to the sum of the masses of the colliding 
PKDGRAV particles. The body was placed at the center of mass and given the center-of-mass speed.} 
Also, we used $R_{\rm tot} = 0.6 R_{\rm Hill}$ and $V_{\rm rand}=0$ in the core runs. The initial rotation
vector of the swarm was set to be aligned with the normal to its heliocentric Keplerian orbit.

The initial radius of PKDGRAV particles, $R$, was set as $R= f r$, where $r$ is the starting boulder size 
and $f$ is an {\it inflation factor} used here to compensate for the fact that the number of PKDGRAV 
particles in the simulation is much smaller than the expected number of bodies in the collapsing swarm.  
Several possible choices of $f$ exist. If PKDGRAV particles are required 
to mimic the actual collision rate of radius $r$ boulders (case A), then  $f^2=n/N=(R_{\rm eq}/r)^3/N$ with 
the initial number of boulders, $n$, being set by the mass constraint. This choice poses problems 
during the late simulation stages, however, because $f=3\times10^6$ with $R_{\rm eq}=250$ km and 
$r=25$ cm. Thus, if $f$ remains constant during the simulation, and a fraction of PKDGRAV particles accrete 
into a body with mass equivalent to, say, a 50-km-radius KBO, the corresponding PKDGRAV particle would have 
radius $R=50 f^{1/3}\approx7,200$~km! This is obviously bad because the separation of components in many known 
binary systems is $<$10,000 km (Noll et al. 2008a).

A different choice of $f$ would be to use $R=R^* = R_{\rm eq}/N^{1/3}$ (case B), which for the above used example 
implies the initial radius $R = 5.4$ km and $\rho=1$ g cm$^{-3}$ of PKDGRAV particles. This setup severely 
underestimates the rate of collisions in the collapsing swarm of real sub-meter boulders, but has the advantage 
that the late stages of accretion of large objects are treated more realistically, because the corresponding 
PKDGRAV particles have adequate radii and bulk densities. 

We conducted simulations with the two extreme setups A and B discussed above, and 
also for several intermediate cases. We define these cases by the initial ratio $f^*=R/R^*$, where $f^*=1$ corresponds 
to case B and $f^*=(n/N)^{1/6}$ to case A. The intermediate cases with $1<f^*<(n/N)^{1/6}$ are probably 
more realistic than the two extreme cases. They conservatively use lower-than-realistic collision rates and 
do not allow the large objects to grow beyond reasonable limits in radius. Specifically, we used $f^*=1$, 
3, 10, 30 and 100.   

Thus, with 3 values of $R_{\rm eq}$, 4 values of $\Omega$ and 5 values of $f^*$, we have 60 different initial 
states of the swarm. Four simulations were performed for each state where different random seeds 
were used to generate the initial positions of PKDGRAV particles in the swarm. We used a 0.3 day 
timestep in the PKDGRAV integrator so that the expected binary orbital periods were resolved by at least
$\sim100$ timesteps. 
We verified that shorter timesteps lead to results similar to those obtained with the 0.3 day timestep. 
The integration time was set to $T_{\rm int} = 100$ yr, or about $0.6 P(30)$ where $P(30)$ 
is the orbital period at 30 AU. Together, our core simulations represent 240 jobs each requiring about 2 weeks 
on one Opteron 2360 CPU. To increase the statistics in the most interesting cases, 10 simulations with different 
random seeds were performed for $\Omega=0.75 \Omega_{\rm circ}$, $f^*=10$ and all $R_{\rm eq}$ values.

Our second suite of simulations includes a diverse set of jobs in which we tested a broader range of parameters, 
extended selected integrations over several orbital periods at 30 AU, used different $R_{\rm tot}$ and 
$V_{\rm rand}$ values, included effects of inelastic bouncing of PKDGRAV particles, imposed retrograde 
rotation of the initial swarm, etc. We describe the results of these simulations in section 3.

\section{Results}

While our core simulations with $f^*>30$ produce massive bodies that are frequently bound in binary systems, the 
binary separations tend to be very large because the inflated PKDGRAV particles prevent formation of tight binaries.
On the other hand, the simulations with $f^* < 3$ show low collision rates and do not produce massive objects in 
100 yr. Moreover, as expected, simulations with $\Omega > \Omega_{\rm circ}$ lead to the swarm's dispersal due to 
excess angular momentum. We therefore first discuss the results obtained with intermediate values of $f^*$, which are 
probably the most realistic ones, and $\Omega \leq \Omega_{\rm circ}$. All binary systems produced in these simulations 
were followed for 10,000 yr to check on their stability and orbital behavior. 

The binary systems that form in $T_{\rm int}=100$ yr are usually complex, typically including two or more large objects and
hundreds of smaller bodies. Over the next 10,000 yr, these systems clear out by collisions and dynamical 
instabilities. In all cases analyzed here the final systems are remarkably simple. They typically include 
a binary with two large objects, and one or two small satellites on outer orbits with separations 
exceeding by a factor of a few the separation of the inner pair. We have not followed these systems for longer 
timespans. It is likely that most of the small, loosely bound satellites would not survive Gyr of dynamical 
and collisional evolution in the KB (Petit \& Mousis 2004).

\begin{figure}
\epsscale{0.8}
\plotone{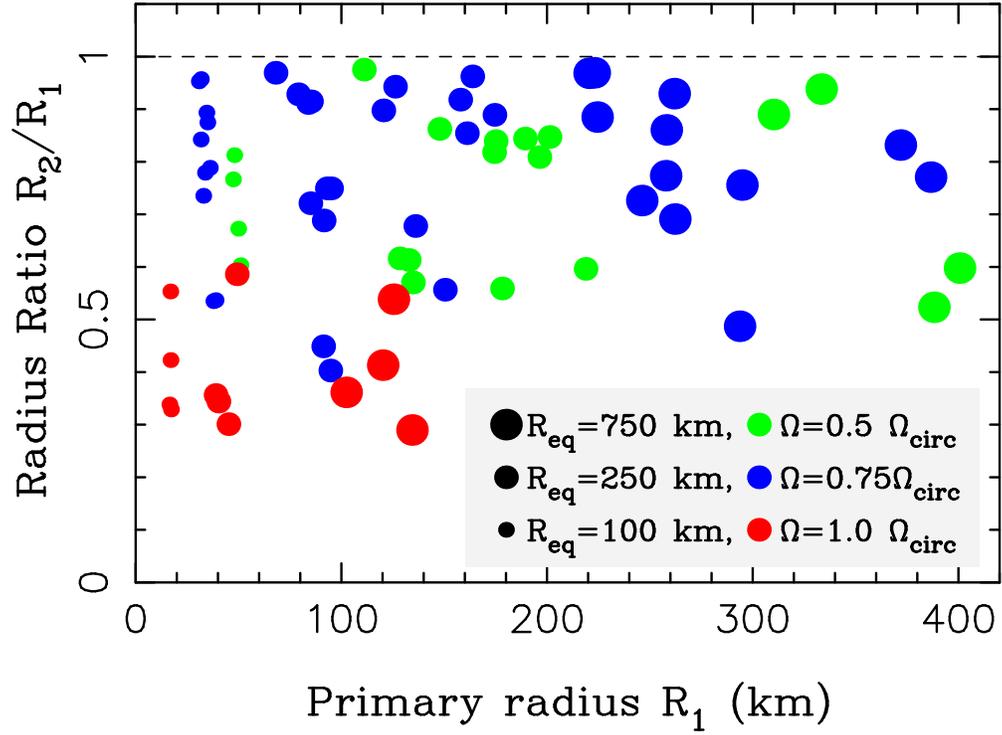}
\caption{The primary radius, $R_1$, and secondary-to-primary radius ratio, $R_2/R_1$, for the binary
systems obtained in our core simulations. The different sizes of symbols correspond to the results
obtained with different initial masses of the collapsing swarm (see legend). Colors indicate the $\Omega$ 
values: $\Omega=0.5 \Omega_{\rm circ}$ (green), $\Omega=0.75 \Omega_{\rm circ}$ (blue),
and $\Omega=\Omega_{\rm circ}$ (red).}
\label{ratio}
\end{figure}

Figure \ref{ratio} shows the primary radius, $R_1$, and the secondary over primary radius ratio, $R_{\rm 2}/R_{\rm 
1}$, obtained for binaries that formed in the runs 
with intermediate values of $f^*$. Each of these simulations, done for different $R_{\rm eq}$, 
$\Omega \leq \Omega_{\rm circ}$ and random seeds, produced at least one binary with similar-size large components. 
In some cases, more than one separate binary systems were found. Values of $R_{\rm 2}/R_{\rm 1}$ obtained here range 
between $\sim$0.3 and 1 with most systems having $R_{\rm 2}/R_{\rm 1}>0.7$. For example, if we limit the statistics 
to $\Omega < \Omega_{\rm circ}$, about 80\% of binary systems have $R_{\rm 2}/R_{\rm 1}>0.7$.

\begin{figure}
\epsscale{0.8}
\plotone{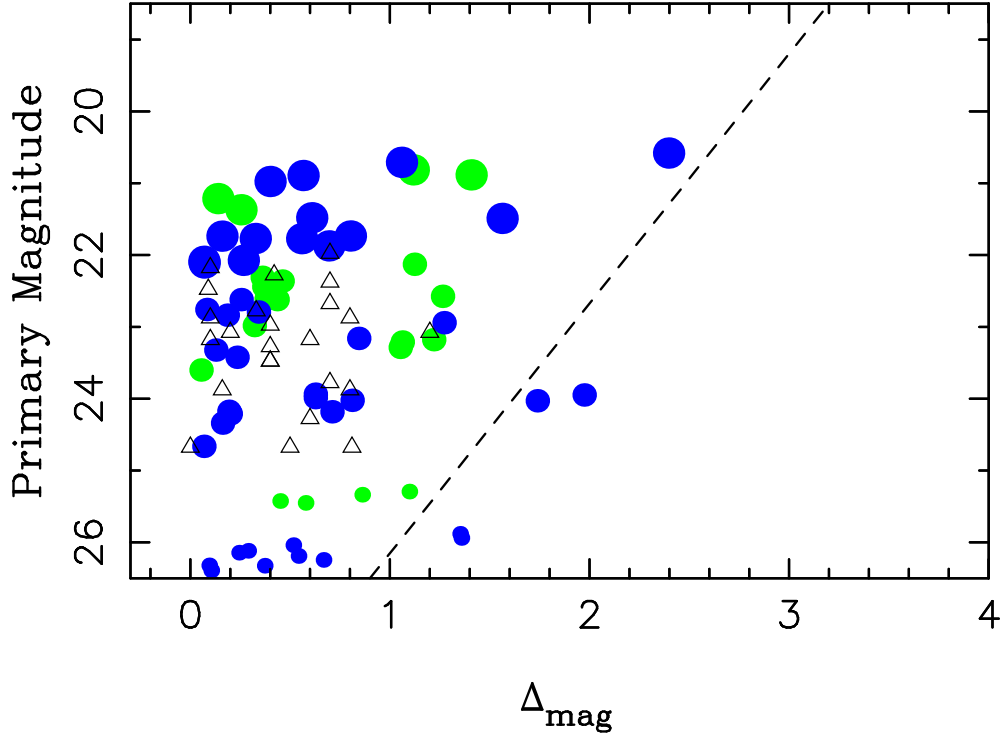}
\caption{Comparison of simulated (color symbols) and observed binaries (triangles). The y-axis shows the 
apparent magnitude of the primary component. The x-axis shows the range of magnitude differences, 
$\Delta_{\rm mag}$. The dashed line corresponds to an approximate empirical detection limit for objects separated by 3 pixels 
from their primary; i.e., 75 milliarcsec. The background at this separation is dominated by the point spread 
function of the primary to a degree that varies as a function of primary magnitude. The figure includes all known 
binary KBOs in the classical KB (Noll et al. 2008a). The simulated binaries were obtained for $\Omega=0.5 \Omega_{\rm circ}$ 
(green symbols) and $\Omega=0.75 \Omega_{\rm circ}$ (blue) (see legend in Fig. \ref{ratio}).}
\label{mag}
\end{figure}

\begin{figure}
\epsscale{0.8}
\plotone{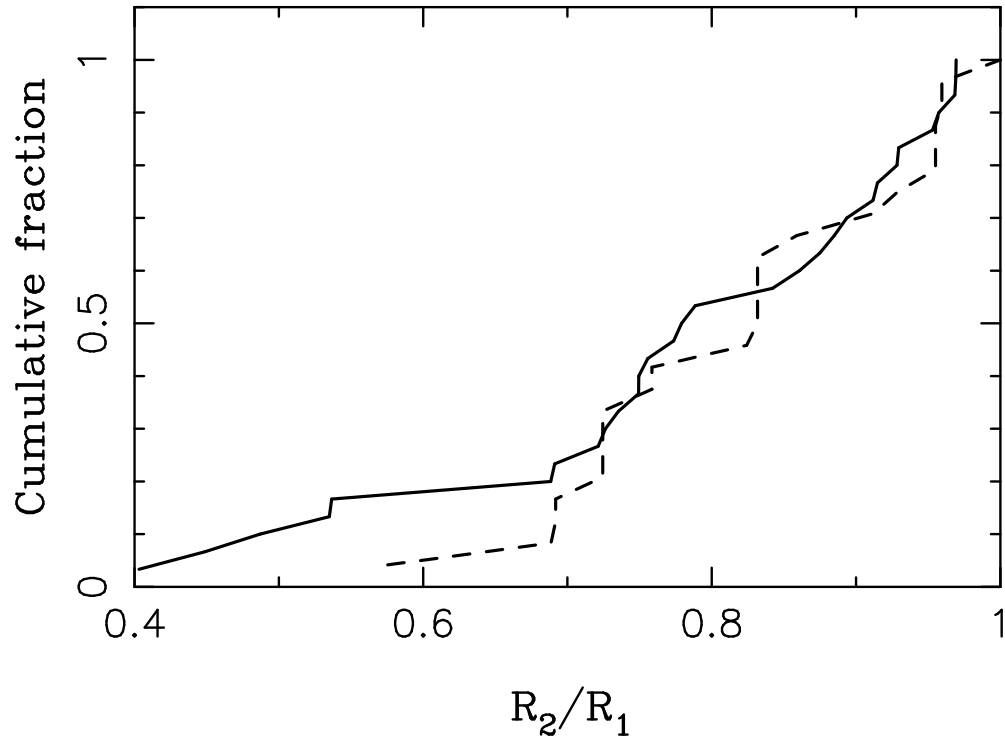}
\caption{The cumulative distribution of $R_2/R_1$ for simulated binaries (solid line) and classical binary 
KBOs (dashed line). The latter were taken from Noll et al. (2008a). We used $\Omega=0.75 \Omega_{\rm circ}$
and $f^*=10$ for this figure. Other values of $\Omega < \Omega_{\rm circ}$ and $f^*\geq3$ lead to a similar 
result.}
\label{cumul}
\end{figure}

We compare our results to observations in Figs. \ref{mag} and \ref{cumul}. Figure \ref{mag}
shows the primary magnitude and magnitude difference, $\Delta_{\rm mag}=5\log_{10}(\sqrt{p_1/p_2}R_1/R_2)$, for 
the simulated binaries and known binary KBOs in the classical KB ($p_1$ and $p_2$ are the albedos of the binary 
components). We assumed $p_1=p_2=0.08$ and heliocentric distance of 44~AU for the model results. The observed binary 
KBO parameters were taken from Noll et al. (2008a). Most simulated binaries have $\Delta_{\rm mag}<1$, in good agreement 
with observations. The results that match the present observations the best were obtained with $R_{\rm eq} = 250$~km. 

The simulated distribution of $R_2/R_1$ is compared to observations in Fig. \ref{cumul}. The match
is strikingly good given the various uncertainties and approximations in our core simulations, except for 
$R_{\rm 2}/R_{\rm 1}<0.7$, where the number of simulated binaries shows a slight excess. Note that the 
observations are incomplete for small $R_{\rm 2}/R_{\rm 1}$ values, because it is hard to identify 
faint satellites near bright primaries. 

\begin{figure}
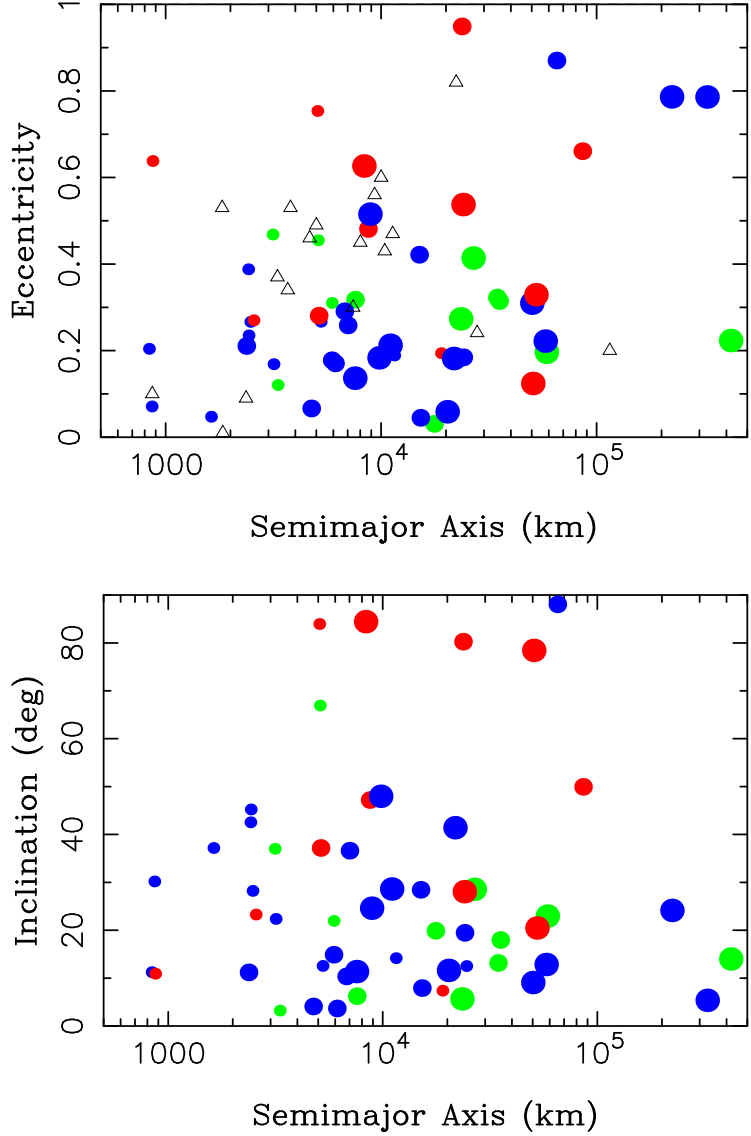

\epsscale{0.6}
\plotone{fig5a.eps}\\[5.mm]
\plotone{fig5b.eps}
\caption{The orbits of simulated binaries: (top) semimajor axis and eccentricity, and (bottom) semimajor axis and 
inclination. The inclination is given with respect to the initial angular momentum vector and does {\it not} represent
the expected distribution of inclination with respect to the Laplace plane. Color symbols show model results for 
different $\Omega$ and $R_{\rm eq}$ values. See legend in Fig. \ref{ratio} for their definition. The triangles 
plotted in the top panel show $a$ and $e$ for all binaries for which we have data (Noll et al. 2008a, Grundy et al. 2009).}
\label{ecc}
\end{figure}

The binary orbits obtained in our simulations are shown in Fig. \ref{ecc}. The semimajor axis values range 
from $\sim$10$^3$ to several $10^5$ km. Most eccentricity values are below 0.6 but cases with $e>0.6$ do also occur.
The observations of binary KBOs show similar trends (Fig. \ref{ecc}(top)). Notably, the orbits of several binary systems 
obtained in the simulations with $R_{\rm eq}=750$ km are similar to that of 2001 QW$_{322}$, which has 
$a\approx120,000$ km and $e\lesssim0.4$ (Petit et al. 2008). This suggests that gravitational collapse can provide
a plausible explanation for the 2001 QW$_{322}$ system. The large orbit of 2001 QW$_{322}$ is difficult 
to explain by other formation mechanisms discussed in section 1.1. 

The binary inclinations show a wide spread about the plane of the angular momentum of the initial swarm 
($\lesssim$50$^\circ$ with only a few cases having $50^\circ<i<90^\circ$). Only one of the simulated binaries was 
found to have switched to retrograde rotation with respect to that of the original swarm. The prograde-to-retrograde 
ratio of binaries produced by GI will therefore mainly depend on the angular momentum vector orientations of 
the collapsing swarms. The normalized angular momentum of the simulated binary systems, $J/J'$ (see, e.g., Noll et al. 
(2008a) for a definition), ranges between $\sim$0.4 and $\sim$5, with larger values occurring for larger separations. 
For comparison, the known binary KBOs in the classical KB have $0.3\lesssim J/J'\lesssim3.5$.  

Interestingly, we do not find any strong correlation between the obtained $J/J'$ values of the final binary systems,
or equivalently their separation, and the assumed initial rotation $\Omega$ of the swarm. Such a correlation would 
be expected if most of the swarm's angular momentum ends up in $J/J'$. The lack of it shows how the angular 
momentum is distributed among the accreting bodies. If there is too much momentum initially 
($\Omega \sim \Omega_{\rm circ}$), only a relatively small fraction of the mass and momentum ends up in the final 
binary. Indeed, it is clear that much mass is lost in the $\Omega = \Omega_{\rm circ}$ case as both $R_1$ and 
$R_2/R_1$ are on the low end of the distribution (Fig. \ref{ratio}).

We found that several stable {\it triple} systems were produced in the simulations. For example, one of the simulations with
$\Omega = 0.75 \Omega_{\rm circ}$ produced a triple system with $R_1=126$~km, $R_2=119$ km and $R_3=77$ km,
where $R_3$ denotes the radius of the smallest component on the outer orbit. For comparison, (47171) 1999 TC$_{36}$ has 
$R_1=140$ km, $R_2=129$ km and $R_3=67$ km (Benecchi et al. 2010). The two orbits of the simulated triple 
are nearly coplanar ($\Delta i$ = 5$^\circ$) and have low eccentricities (0.2 and 0.3, respectively). These 
properties are again reminiscent of (47171) 1999 TC$_{36}$. The separations of components in the simulated triples, 
including the one discussed here, tend to be a factor of a few larger than those in (47171) 1999 TC$_{36}$ 
(867 and 7411 km, respectively; Benecchi et al. 2010).   

\begin{figure}
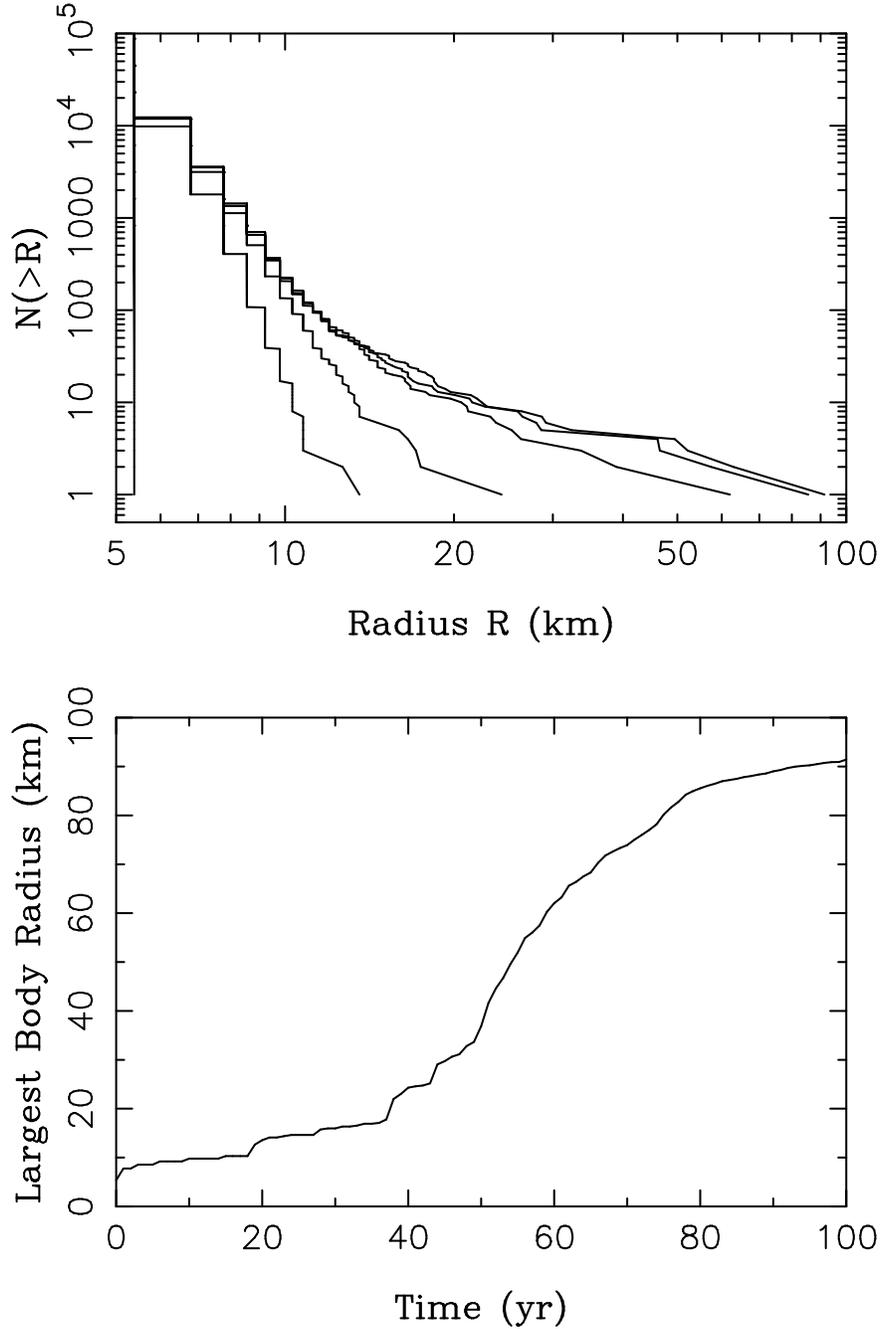

\epsscale{0.7}
\plotone{fig6a.eps}\\[5.mm]
\plotone{fig6b.eps}
\caption{The cumulative size distribution (top) and radius of the largest body (bottom) in the simulation
with $R_{\rm tot}=0.6 R_{\rm Hill}$, $R_{\rm eq}=250$ km, $\Omega=0.75\Omega_{\rm circ}$ and $f^*=10$. The top panel 
shows six snapshots of the cumulative size distribution, $N(>R)$, spaced by 20 yr in time from $t=0$ to 100 yr. 
The size distribution curves rise and move from left to right with $t$ as large objects accrete in the swarm.}
\label{sfd1}
\end{figure}

Figure \ref{sfd1} illustrates the size distribution of bodies growing in the collapsing swarm for $R_{\rm eq}=250$ km, 
$\Omega=0.75\Omega_{\rm circ}$ and $f^*=10$. Initially, bodies grow by normal accretion for which the growth rate 
of an object is not a strong function of its mass. Upon reaching a threshold of $R\sim20$ km, however, the largest 
objects start growing much faster than the smaller ones. This is diagnostic of runaway growth (see, e.g.,  
Kortenkamp et al. 2000 for a review). Runaway growth occurs in the collapsing swarm because the collisional 
cross-section of the largest bodies is strongly enhanced by gravitational focusing. 

\begin{figure}
\epsscale{0.8}
\plotone{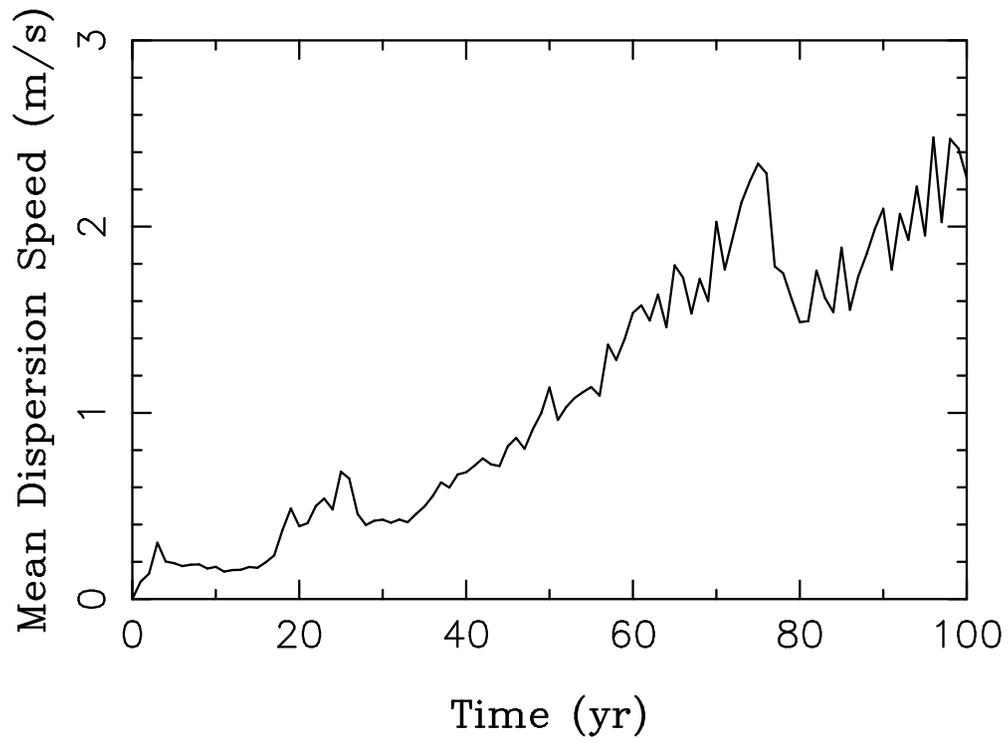}
\caption{Mean dispersion speed of particles in the simulation with $R_{\rm eq}=250$~km, $\Omega=0.75\Omega_{\rm circ}$ 
and $f^*=10$.}
\label{dispv}
\end{figure}

Figure \ref{dispv} shows the mean dispersion speed, $V_{\rm disp}$, 
of bodies in the collapsing swarm as a function of time. It slowly increases due to dynamical stirring from large 
bodies but stays relatively low during the whole simulation ($V_{\rm disp}\lesssim2$ m s$^{-1}$). 
This leads to a situation in which the escape speed, $V_{\rm esc}$, of $R > 10$ km 
bodies largely exceeds $V_{\rm disp}$, and the runaway accretion begins. Note also that the size distribution 
does not change much after 80 yr, because the large bodies run out of supply. This shows that the integration 
timespan was roughly adequate in this case. 
 
We now turn our attention to the results obtained with $f^*=1$. Our core simulations with $f^*=1$ show little 
accretion because the collisional cross-section of PKDGRAV particles is small in this case. This suggests
that a longer integration timespan is needed for $f^*=1$. We extended several core integrations with $f^*=1$ to 
$T_{\rm int}=1000$ yr, or about 6 orbital periods at 30 AU, and found that large objects accrete in these 
extended simulations in very much the same way as illustrated in Fig. \ref{sfd1}. The binary properties obtained 
in the extended runs with $f^*=1$ were similar to those discussed above, but better statistics will be needed 
to compare the results more carefully.

In additional tests, we used the same $M_{\rm tot}$ as in the core simulations and $R_{\rm tot}=0.4 R_{\rm Hill}$ 
to see how things would work for a very dense initial concentration of solids. With $f^*=10$ we found that
the largest object that grows out of the swarm has $R=150$ km (compared to $R=92$ km for $R_{\rm tot}=0.6 R_{\rm Hill}$). 
Notably, large bodies can also rapidly form with $f^*=1$ in this case, the largest having $R=110$ km (compared 
to $R=22$ km for $R_{\rm tot}=0.6 R_{\rm Hill}$). On the other hand, simulations with $R_{\rm tot}=0.8 R_{\rm Hill}$
lead to smaller $R$ values, probably because the shearing effects become important when $R_{\rm tot}$ approaches
$R_{\rm Hill}$.

This shows that the accretion timescale sensitively depends on the initial concentration of solids in the 
collapsing cloud. For a reference, with $R_{\rm tot}=R_{\rm Hill}$ at 30 AU we obtain a 
concentration of solids, $\rho_{\rm solids}$, about 15 times greater than that of the gas in the standard Minimum 
Mass Solar Nebula ($\rho_{\rm gas}$; Hayashi et al. 1981), while $R_{\rm tot}=0.4R_{\rm Hill}$ leads to 
$\rho_{\rm solids}/\rho_{\rm gas}\sim230$. These values are in the ballpark of the ones produced in the simulations 
of Johansen et al. (2009) for protoplanetary disks with slightly enhanced metallicity. 

We also performed several additional simulations with $V_{\rm rand}\neq0$ and/or inelastic bouncing\footnote{In this 
approximation, the colliding PKDGRAV particles merge only if their impact speed is below a specific threshold. 
Otherwise the particles bounce with a loss of energy parameterized by the normal coefficient of restitution. 
Sliding friction was ignored in our simulations.} of PKDGRAV particles. These tests showed that binary formation 
occurs over a broad range of $V_{\rm rand}$ and restitution coefficient values, so long as the initial $V_{\rm rand}$ value 
is significantly smaller than $V_{\rm circ}$. Placing a hard upper limit on $V_{\rm rand}$ as a function of other 
parameters, however, will require a systematic sampling of parameter space that is beyond the scope of this paper. 

\section{Discussion}

We found that the observed propensity for binary Kuiper Belt Objects (KBOs) and their properties can be a natural 
consequence of KBO formation by Gravitational Instability (GI). The binary formation in GI is robust, directly 
linked to the formation of large KBOs, and does not require finely tuned size distributions invoked by the HC models 
(see, e.g., Noll et al. 2008a). The common colors of the components of binary KBOs, their orbital parameters, 
including the wide binary systems such as 2001 QW$_{322}$,
and triple systems such as (47171) 1999 TC$_{36}$, can be readily explained in this context. Moreover, the binary 
fraction in the KB expected in the GI model is large reaching $\sim$100\% for a broad range of initial parameters.
This favorably compares with observations that indicate, when extrapolated to smaller binary separations, that 
$>$50\% of classical low-$i$ KBOs are binary systems (Noll et al. 2008a). 

The inclination distribution of binary orbits can help to constrain KB formation (Schlichting \& Sari 2008b). 
Unfortunately, the binary orbits determined so far typically have a pair of degenerate solutions representing 
reflections in the sky plane. These solutions have the same $a$ and $e$, but different inclinations. The very 
few unique inclination solutions that have been reported up to now seem to indicate that the binary orbits can be 
prograde ($i<90^\circ$, (42355) Typhon/Echidna; Grundy et al. 2008), retrograde ($i>90^\circ$, 2001 QW$_{322}$; Petit 
et al. 2008) or nearly polar ($i\sim90^\circ$, (134860) 2000 OJ$_{67}$ and 2004 PB$_{108}$; Grundy et al. 2009). 

The broad distribution of binary inclinations should be a signature of the formation mechanism rather than that of 
the later evolution because the long-term dynamical effects should not have a strong impact on the binary orbits 
with $i<40^\circ$ and $i>140^\circ$, 
and cannot switch from prograde to retrograde motion (or vice versa; Perets \& Naoz 2009). To explain the retrograde 
orbits in the GI model, we thus probably need to invoke a retrograde rotation of the collapsing clump, while the 
simulations of Johansen et al. (2007, 2009) seem to generally indicate prograde rotation. This issue needs to be studied 
in a more detail, however, using a better resolution in the dynamical codes. The rotation direction of clumps in the 
model of Cuzzi et al. (2008) is uncertain. 

Our binary formation model could also potentially apply to the Orcus-Vanth and Pluto-Charon systems, although the 
corresponding large $M_{\rm tot}$ values were not studied here. 

Observations by Brown et al. (2010) imply sizes of Orcus and Vanth of 900 and 280 km, respectively, a mass 
ratio of $33$, if equal densities and albedos are 
assumed, and the semimajor axis of the binary orbit $8980 \pm 20$ km. This mass ratio and orbit would be consistent 
with formation from a giant impact and subsequent outward tidal evolution of the binary orbit. Assuming a factor of 
2 lower albedo for the non-icy Vanth, however, implies sizes of 820 and 640 km and a mass ratio of 2 (Brown et al. 
2010). Such parameters could be difficult to reconcile with the impact formation of the Orcus-Vanth system and could 
rather indicate a different formation mechanism, perhaps akin to that studied in this work. Physical properties of the 
Orcus-Vanth system need to be determined better to discriminate between different formation models.

Using impact simulations, Canup (2005) was able to 
explain the main properties of the Pluto-Charon system (e.g., $\sim$15\% mass ratio, $J/J'\sim 0.4$) using an oblique, 
low-speed impact of an object that contained 30-50\% of the current Pluto-Charon mass. It remains to be shown, however, 
whether such collisions were sufficiently common in the early KB since the relevant timescale could be long (Canup 
2005). On the other hand, formation of the Pluto-Charon system by gravitational collapse would require very large 
$M_{\rm tot}$ of the collapsing swarm, which can be a challenge for the GI theories. Interestingly, a hybrid formation 
model (collapse followed by an impact) is also possible, because low-speed collisions between large bodies commonly 
occur in our simulations.

Note that the precursor binary system similar to Pluto-Charon is needed to explain the capture of Neptune's moon 
Triton by exchange reaction (Agnor \& Hamilton 2006), indicating that these massive binary systems were once 
common in the outer solar system.  

Wide binary systems with similar-size components could have also formed in the inner solar system. Indeed,
constraints from the Size Frequency Distribution (SFD) of main-belt asteroids indicate that the standard 
hierarchical coagulation was {\it not} the driving force of planetesimal accretion at 2-4~AU (Morbidelli 
et al. 2009). Instead, asteroids have probably formed by the GI-related processes (Johansen et al. 2007, Cuzzi 
et al. 2010). The results of gravitational collapse simulations presented here, when scaled to a smaller Hill
radius at 2-4 AU, can therefore also be applied to the asteroid belt. If so, it may seem puzzling why wide binaries 
with similar-size components are not detected in the asteroid belt today.

We speculate that wide asteroidal binaries, if they actually formed, would have been disrupted by collisions and 
scattering events during the subsequent evolution. For example, even a relatively small impact on one of the two 
binary components can impart enough momentum into the component's orbit to unbind it from its companion. This can  
happen when roughly $m_i v_i > m_b v_b$, where $m_i$ and $v_i$ are the mass and speed of the impactor, and $m_b$ 
and $v_b$ are the mass and speed of the binary component (see Petit \& Mousis 2004). For the component 
radii $R_1=R_2=50$ km, density $\rho=2$ g cm$^{-3}$, $v_b \sim 10$~m~s$^{-1}$ (corresponding to separation 
$\sim$0.05 $R_{\rm Hill}\approx1000$ km at 2.5 AU), and $v_i=5.8$~km~s$^{-1}$ (Farinella \& Davis 1992), 
this would imply the impactor mass $m_i \gtrsim 10^{-3} m_b$ or, equivalently, impactor radius $r_i 
\gtrsim 5$ km (for $\rho=2$ g cm$^{-3}$) for the binary to become unbound.

Since, according to Bottke et al. (2005), there are $N_i \sim 10^4$ asteroids with $r_i > 5$ km in the present 
asteroid belt, we can estimate that the {\it present} rate of unbinding collisions would be $\sim 2 P_i N_i R_1^2 = 
2 \times 10^{-10}$ yr$^{-1}$, where  $P_i=2.8\times10^{-18}$ km$^{-2}$ yr$^{-1}$ is the intrinsic collision 
probability (see, e.g., Farinella \& Davis (1992) for a definition), and the factor of two appears because the impact 
can happen on any of the two components. This would indicate binary lifetimes comparable to the age of the solar 
system. The number of relevant impactors $N_i$, however, was likely much larger in the past, perhaps by a factor of 
10-1000 (Weidenschilling 1977, Petit et al. 2001, Levison et al. 2009), than in the present asteroid belt. In addition, 
gravitational scattering from large planetary embryos, thought to have formed in the main-belt region (Petit et al. 
2001), would have also contributed to disruption of wide binaries. It thus seems unlikely that a significant 
fraction of wide asteroid binaries could have survived to the present times.

A notable exception of an asteroid binary produced by gravitational collapse may be (90) Antiope (Merline et al. 
2000), which is the only known asteroid binary with large, equal-size components ($R_1 \sim R_2 \sim 45$ km). We speculate 
that the small separation of components in the Antiope system (only $\sim$170 km) could have been a result of the tidal 
evolution of the original, possibly much wider orbit. Indeed, it has been pointed out that wide binaries with orbits 
that are significantly inclined (inclinations $39.2^\circ<i<140.8^\circ$) undergo Kozai oscillations 
during which the tidal dissipation is especially effective, and can shrink and circularize the binary orbit (Perets \& 
Naoz 2009). For reference, the current inclination of the Antiope's binary orbit is $\sim40^\circ$ (Descamps 
et al. 2009). Alternatively, the (90) Antiope system could have formed by impact-induced fission of a 100-km parent 
asteroid (e.g., Weidenschilling et al. 2001).

The survival of binary KBOs after their formation is an open problem. Petit \& Mousis (2004) have estimated that
several known binary KBOs (e.g., 1998 WW$_{31}$, 2001 QW$_{322}$ and 2000 CF$_{105}$) have lifetimes against collisional 
unbinding that are much shorter than the age of the solar system. These estimates were based on an assumed relatively 
steep SFD extending down to $r_i=5$ km, which favors binary disruption, because of the large number of available 
impactors. When we update Petit \& Mousis' estimates with a probably more reasonable SFD of KBOs given by Fraser et al. 
(2008), which is steep down to 60-95 km and then very shallow (differential power index $\sim1.9$), we find that a 
typical 100-km-class wide binary KBO is unlikely to be disrupted over 4 Gy ($\lesssim1$\% probability), 
except if the KB was much more massive/erosive in the past. This poses important constraints on KB formation 
as it may indicate that the classical low-$i$ KBOs formed in a relatively quiescent, low-mass environment. 

Levison et al. (2008) proposed that most of the complex orbital structure seen in the KB region today (see, e.g., 
Gladman et al. 2008) can be explained if bodies native to 15-35 AU were scattered to $>$35 AU by eccentric Neptune (Tsiganis et al. 
2005). If these outer solar system events coincided in time with the Late Heavy Bombardment (LHB) in the inner solar 
system, as argued by Gomes et al. (2005), binaries populating the original planetesimal disk at 15-35 AU would have 
to withstand $\sim$700 My before being scattered into the Kuiper belt. Even though their survival during this epoch is 
difficult to evaluate, due to major uncertainties in the disk's mass, SFD and radial profile, the near absence of 
binaries among 100-km-sized hot classical KBOs (Noll et al. 2008a,b)  seems to indicate that the unbinding collisions and 
scattering events must have been rather damaging. The (617) Patroclus-Menoetius binary system, thought to have been captured 
into its current Jupiter-Trojan orbit from the 15-35 AU disk (Morbidelli et al. 2005), can be a rare survivor of the 
pre-LHB epoch, apparently because its relatively tight binary orbit ($a=680$ km; Marchis et al. 2006) resisted disruption.

\acknowledgements

We thank Bill Bottke, Hal Levison and Alessandro Morbidelli for stimulating discussions, and an anonymous referee
for a very helpful report on the manuscript.

\appendix
\section{Role of Gas Drag}
While aerodynamic forces are crucial in creating dense clumps, they are less important in the final 
collapse phase. The aerodynamic stopping time of a rock with radius $r$, density $\rho$ and mass 
$m$ is
$$t_{\rm stop} = {\rho r \over \rho_{\rm gas} c_{\rm gas}}\ ,
$$
where $\rho_{\rm gas}$ is the gas density, and $c_{\rm gas}$ is the sound speed.  

For a rough estimate of the collision rate, we assume that the solid mass is distributed in a sphere 
with fractional radius $f_{\rm H}$ of the Hill radius, giving a number density 
$n \sim  (M_{\rm tot}/m)/(f_{\rm H} R_{\rm Hill})^3$ 
and a virial speed $v \sim \sqrt{G M_{\rm tot}/(f_{\rm H} R_{\rm Hill})}$.  
With a geometric cross section, $\sigma \sim r^2$, the collision time $t_{\rm coll} \sim 1/(n \sigma v)$ gives a ratio
$$
{t_{\rm coll} \over t_{\rm stop}} \sim {\Sigma_{\rm gas} a_\odot^2 \over M_\odot} \left(M_\odot 
\over M_{\rm tot} \right)^{1/3} f_{\rm H}^{7/2} \approx  0.05 \sqrt{a_\odot \over 30 {\rm AU}}{250{\rm km} 
\over R_{\rm eq}}f_{\rm H}^{7/2}\ ,
$$
where $a_\odot$ is the distance to the Sun and $\Sigma_{\rm gas} \sim \rho_{\rm gas} c_{\rm gas}/\Omega_{\rm Kep}$ 
is the gas surface density.

We thus estimate that collisions are dominant when collapse begins and $f_{\rm H} \sim 1$.  The strong 
dependence on $f_{\rm H}$ means that collisions become even more dominant as collapse proceeds.

We also estimate that drag forces do not have a strong effect on a binary that forms by collapse.  
The KBO size $R$ now exceeds the gas mean free path and turbulent drag applies with a characteristic 
timescale
$$t_{\rm drag} \sim {\rho R \over \rho_{\rm gas} v_{\rm orb}} \approx 8~{\rm Gyr} \left(a_\odot \over 
30 {\rm AU}\right)^{2.8}\sqrt{{a_{\rm b} \over 10^4{\rm km}}{100{\rm km} \over R}}\ .
$$
For simplicity, we assumed a binary system with equal mass components, circular binary orbit 
with separation $a_{\rm b}$ and orbital speed $v_{\rm orb}$. Since $t_{\rm drag}$ largely exceeds the 
$\sim$ Myr lifetime of the gas disk, the effect of gas drag on the binary orbit is negligible.

\clearpage

\clearpage

\clearpage

\clearpage

\clearpage

\clearpage

\clearpage

\end{document}